\documentclass[aps,pre,a4paper,reprint,notitlepage,superscriptaddress]{revtex4-1}
\usepackage{amsmath,amssymb}
\usepackage[dvipdfmx]{graphicx}
\usepackage{bm}
\usepackage{color}

\begin{document}

\title{Large-deviation Properties of Linear-programming Computational 
Hardness  of the Vertex Cover Problem
}
\date{\today}

\author{Satoshi Takabe}
\email{s\_takabe@nitech.ac.jp}
\affiliation{Department of Computer Science, Nagoya Institute of Technology, Gokiso-cho, Showa-ku, Nagoya, Aichi, 466-8555, Japan}
\author{Koji Hukushima}
\affiliation{Graduate School of Arts and Sciences, The University of Tokyo, 3-8-1 Komaba, Meguro-ku, Tokyo 153-8902, Japan}
\author{Alexander K. Hartmann}
\affiliation{Institut f\"ur Physik, Carl von Ossietzky Universit\"at Oldenburg, 26111 Oldenburg, Germany
}

\begin{abstract}
The distribution of the computational cost of linear-programming (LP) relaxation 
for vertex cover problems on Erd\"os-R\'enyi random graphs
  is evaluated by using the rare-event sampling method.
As a large-deviation property, {differences of the distribution
for ``easy'' and ``hard'' problems are} found 
reflecting the hardness of approximation by LP relaxation.
{In particular}, by evaluating the total variation distance between 
conditional distributions with respect to the hardness,
it is suggested that those distributions are {almost indistinguishable} 
in the  replica symmetric (RS) phase while they asymptotically 
{differ} in the replica symmetry breaking (RSB) phase.
In addition, {we seek for a relation to graph structure by
investigating} a similarity to bipartite graphs,
 which exhibits a quantitative difference between the RS and RSB phase.
These results indicate the nontrivial relation of the typical 
computational cost of LP relaxation to the RS-RSB phase transition 
{as present in the spin-glass theory of models
on the corresponding} random graph structure.
\end{abstract}


\maketitle
\section{Introduction}
{In statistical mechanics, combinatorial optimization problems have
attracted a great deal of attention during the past two decades
\cite{phase-transitions2005,26,moore2011}.  The motivation and hope is to obtain an
understanding of the source of computational hardness 
which is inherent
to many combinatorial optimization problems \cite{papadimitriou1998}. 
The statistical mechanics approaches are based on taking a physics perspective
and applying notions and techniques from the the field of disordered
systems like spin glasses. Actually, from the computer science point of view,
}
understanding the computational hardness is still lacking.
This is visible from the fact that the famous P--NP problem \cite{garey1979}
 is still not solved.
Here, P is a class of ``easy'' problems that can be solved {in the worst
case} in polynomial time
and NP contains the problems for which the solution can {only}
be  checked in  polynomial time.
So far, for no problem in NP, an algorithm is known which finds solutions 
of the problems in worst case in polynomial time.
On the other hand, it is not proved that no such algorithm will exist.

In contrast to the worst-case analysis, one is often interested in the 
typical case running time, {in particular for real-world
applications. 
From a fundamental science point of view,}
by studying the properties of {given} easy and hard samples, 
one might be able to understand the source of computational complexity.
For this reason, in statistical mechanics one started to investigate 
{random ensembles of NP-hard problems, which are typically
controlled by one or several parameters. Examples are 
satisfiability problems with random formulas, 
the traveling salesperson problems for randomly distributed cities, or the
vertex-cover {(VC)} problems on random graphs. For these ensembles
phase transitions have been observed at the boundary between a phase
where the random instances are typically easy to solve and another phase
where
the random instances are typically hard.
\cite{mitchell1992,monasson1997,mertens1998,wh1,phase-transitions2005}.
}

{Interestingly, the actually hardest instances are often found
right \emph{at} the phase transitions. This reminds the statistical physicist on
the \emph{critical slowing down} which appears, e.g., for the Monte Carlo
simulation of the ferromagnetic Ising model.
Thus, to understand the source of computational hardness best, it might be
beneficial to study the hardest instances available, e.g.,
the problem instances which are located right at the
phase transitions. As an alternative approach, some attempts have been performed
to construct hard instances guided by physical insight 
\cite{horie1996,hoos2000,liu2015,jia2007}.
Examples are instances designed such that
equivalent spin-glass instances exhibit no bias in the distributions of 
local fields and exhibit a first order transition with a 
backbone \cite{satsat2002} or such that, due to the degeneracy, 
many existing solutions of an optimization problem are isolated 
\cite{zdeborova2006}.
Recently, a heuristic algorithm has been implemented \cite{marshall2016} to 
automatically construct spin-glass instances with respect to
finding the ground state as hard as possible (for
a given ground-state algorithm). We extend this approach
in the present work by sampling instances of an optimization problem
in equilibrium, the hardness  (i.e., the number
of steps of the optimization algorithm) is interpreted as
the ``energy'' of the instance. The sampling is controlled by an artificial
temperature.
 In particular our approach allows us, in addition to obtain very hard
instances,  to obtain the \emph{distribution} of the hardness  over the random 
problem ensemble, even down to the low-probability tails. 
One tail will contain very hard instances, the other tail very easy instances.
Such large-deviation
approaches have been applied, e.g., to study the distribution of scores
of random-sequence alignment \cite{AKH2002}, 
the distribution of the size of the largest
components of random graphs \cite{AKH2011}, 
the distribution of endpoints of fractional
Brownian motion \cite{fBm_MC2013}, 
or the distribution of non-equilibrium work \cite{work_ising2014}. 
This fairly general approach
enables us 
to investigate how the shape of the hardness distribution 
changes
as a function of the control parameter. In particular we can investigate
how the distribution differs in the phases where the
typically easy and the typically hard instances are found, respectively.
}

{In particular,} 
we study the {VC problem} on the Erd\H{o}s-R\'enyi 
random graphs \cite{erdoes1960} with average connectivity $c$ as a {control} parameter.
Here, the easy-hard transition was observed \cite{wh1} 
at $c=e$ ($e$ is the Euler's number).
This transition was independent of the algorithm used, 
branch-and-bound with leaf removal~\cite{Tak} or linear programming {(LP)} with 
cutting-plane approach \cite{dh}.
Analytically, the replica symmetry {(RS)} is broken at $c=e$ as well \cite{wh1}.
These past studies have been performed by studying typical  
instances for different values of $c$.
To our knowledge, nobody has studied full distribution of the 
``computational hardness,'' measured in a suitable way, for this problem.
Also for the other NP-hard problems we are not aware of such studies.
{Here, we use a linear-programming relaxation to solve VC.}
Nevertheless, only obtaining the full distribution of the 
quantities of interest allows the full understanding of any random problem. 
In particular, the tails of the distribution contain the very easy and very 
hard instances.
Studying the properties of those {hardest possible} instances 
might allow to get better understanding of the source of the computational hardness.

The reminder of the paper is organized as follows: Next, we define the 
{VC} problem formally and state the linear-programming algorithm
{which} we applied for solving the 
VC problem. In the third section,
we explain the large deviation approach used to obtain the distribution of hardness.
In section four we present our results. We close the paper by a summary and a 
discussion of our results.

\section{Vertex covers and LP relaxation}
In this section, we define {minimum vertex-cover (min-VC)} problems and introduce LP relaxation as an approximation scheme.

Let $G=(V,E)$ be an undirected graph with a vertex set $V=\{1,\dots,N\}$ and an edge set $E\subset V^2$.
It is assumed that $G$ does not have multi-edges nor self-loops.
A set $S\subset V$ is called vertex cover of $G$ if any edge connects to at least a vertex in $S$.
A vertex in vertex cover $S$ is called covered.
The (unweighted) min-VC problems is then defined as a computational problem to search a vertex cover with the minimum number of covered vertices.
For a mathematical formulation,
we set a binary variable $x_i$ to each vertex $i\in V$, which takes one if vertex $i$ is covered and zero otherwise.
The problem is then represented as an integer programming (IP) problem as follows:
\begin{equation}
\begin{array}{lll}
\mbox{Min.} &\displaystyle\sum_{i=1}^Nx_i, &\\
\mbox{subject to\ } &x_i+x_j\ge 1 &\quad \forall (i,j)\in E, \\
&x_i\in \{0,1\} &\quad \forall i\in V.
\end{array}
\label{eq_v1}
\end{equation}

Min-VCs belong to a class of NP-hard \cite{garey1979}
indicating that no exact algorithm {is known that}
can solve {the problem} {in the worst case in polynomial time 
as a function} of problem size $N$. To avoid this computational 
hardness, approximation algorithms are commonly used.
Especially, the linear programming  relaxation is a fundamental 
approximation scheme for the IP problems.
For LP relaxation, constraints on binary variables, $x_i\in \{0,1\}$, 
are replaced to continuous constraints with interval $[0,1]$.
Thus, the LP-relaxed min-VC problem is represented by
\begin{equation}
\begin{array}{lll}
\mbox{Min.} &\displaystyle\sum_{i=1}^Nx_i, &\\
\mbox{subject to\ } &x_i+x_j\ge 1 &\quad \forall (i,j)\in E, \\
&x_i\in [0,1] &\quad \forall i\in V.
\end{array}
\label{eq_v2}
\end{equation}
It is a kind of LP problems which can be exactly solved in polynomial time
using an ellipsoid method~\cite{Kha}. 
However, since it is computationally costly, other exact algorithms such as Dantzig's simplex method 
and the interior-point method~\cite{Kar} are often used.
Their difference lies in the strategy to search an optimal solution on a facet of the simplex defined by constraints in
Eq.~(\ref{eq_v2}); 
the former moves from an extreme point to another extreme point at each step but the latter can search feasible solutions inside the simplex.
{Note that if a solution of the relaxed problem contains
only integer variables, it is, by minimality, automatically a solution of the
corresponding IP problem.} 
The fact enables us to naively define the hardness of the min-VC problems 
for LP relaxation
 by counting the number of {non-integer values}
 in an LP-relaxed solution.
It is worth noting that {VC} problems have a good property called half integrality;
considering extreme-point solutions of the problem, they are represented by vectors with half integers, $\{0,1/2,1\}$~\cite{nt}. {Thus, for VC,
all non-integer values are $1/2$.}

One of our goals in this paper is to investigate the relationship between the 
structure of {min-VCs on random graphs} and 
its statistical-mechanical picture by using LP relaxation.
To achieve the goal, we mainly examine the computational cost of the simplex method
hopefully reflecting the structure of the simplex given by Eq.~(\ref{eq_v2}).
Since it restricts candidates of optimal solutions 
to feasible solutions lying on extreme points of the polytope,
the computational cost, i.e., the number of iterations reaching to the optimal solution, is regarded as 
a measure of complexity of the simplex {solver} and of the problem.
For the simplex method, there are some solvers and a number 
of initialization schemes and pivot rules
which may change the computational cost {to treat a
 given instanced by the algorithm, respectively.}
As described later, however, our main findings are independent of 
those selections.

Before closing this section, we describe 
{how the random graphs are defined for which we obtain min-VCs.}
In this paper, we examine Erd\H{o}s-R\'enyi (ER) 
random graphs as a random ensemble. 
{Each instance of such a} random graph is generated by,
{starting with an empty graph, 
creating an ?undirected edge for} each pair of vertices with 
probability $c/(N-1)$, {i.e., connecting the two nodes.  Here} 
$c=O(1)$ represents average degree.
Then, the degree distribution converges to the Poisson distribution with mean $c$ as the number of vertices $N$ grows.
As described in the {previous} section, it has been revealed 
that the randomized problems exhibit {a phase transition related to RS and its breaking (RSB)}
at $c^\ast=e$~\cite{wh1},
which is also the threshold where LP relaxation fails to approximate the problems with high accuracy~\cite{dh,th2}.
While the relation is usually observed for other ensembles~\cite{th4},
the ER random graph is the simplest example to investigate the graph structure making the problem hard for LP relaxation.

\section{Rare-event sampling}
As we consider {min-VCs on an ensemble of random graphs}, 
it is necessary to estimate  distributions of observables over those instances.
The rare-event sampling is useful to estimate {distributions
over a large range of the support even into the tails} because it 
significantly reduces the number of samples {needed} compared to 
a simple sampling. {Depending on the application, probabilities
as small as $10^{-20}$ or smaller are easily accessible.}
In computational physics, the task is crucial for various fields and some sampling schemes 
 such as the Wang-Landau method~\cite{WL} and multicanonical method~\cite{berg,IH} have been developed, {which were originally used
to sample configurations with extreme energies in the Boltzmann
ensemble. Nevertheless, instead of the energy the sampling can be
with respect to any measurable quantity~\cite{AKH2002}.
This and similar approaches are, e.g.,} widely used for estimating large-deviation properties
 of {random graph structures~\cite{AKH2015,AKH2017,AKH2017_2}}.

Now we briefly describe the outline of the estimation.
In this paper, we estimate the distribution of number of iterations 
$S$ of an {simplex} solver for LP-relaxed min-VCs for {ER random graphs}.
Since $S$ is considered as a function of a graph $G$, we need to sample graphs following the weight of the ER random graphs with the number of vertices $N$ fixed.
For an unbiased sampling, each graph $G=(V,E)$ follows the distribution given by
\begin{equation}
q_0(G)=(1-p)^{M-|E|}p^{|E|}\quad (G\in\mathcal{G}_N),   \label{eq_c4_b2a3}
\end{equation}
where $p=c/(N-1)$, $M=\binom{N}{2}$, and $\mathcal{G}_N$ is a set of graphs with cardinality $N$.
A biased sampling with respect to $S$ is then executed by sampling graphs with the following distribution:
\begin{equation}
q_T(G)=Z(T)^{-1}q_0(G)e^{-S(G)/T}\quad (G\in\mathcal{G}_N),\label{eq_c4_b3a}
\end{equation}
where $T$ represents a ``temperature'' of the system, {which can
also be negative,}
and the partition function $Z(T)$ is given by
\begin{equation}
Z(T)= \sum_{G\in\mathcal{G}_N} q_0(G)e^{-S(G)/T}. \label{eq_c4_b2a4}
\end{equation}
{In practical simulations \cite{practical_guide2015}, the choice of the temperature
determines the range of the values of $S$ where the sampling of
graphs is concentrated.}
Please notice that the unbiased sampling corresponds to the {$T\rightarrow\pm\infty$ limit} in Eq.~(\ref{eq_c4_b3a}).

The reweighting method is used to {obtain the distribution of $S$
over a large range of the support. The distribution} 
reads
\begin{equation}
P(S)=\sum_G \delta_{S,S(G)}q_0(G), \label{eq_c4_b6i}
\end{equation}
where $\delta_{n,m}$ represents a Kronecker delta.
Biased  distribution $P_T(S)$ with temperature $T$ is represented by
\begin{align}
P_T(S)&\equiv {Z(T)}^{-1}e^{-S/T}\sum_{G\in\mathcal{G}_N} \delta_{S,S(G)}q_0(G)\nonumber\\
&={Z(T)}^{-1}e^{-S/T}P(S). \label{eq_c4_b5}
\end{align}
This indicates that the unbiased  distribution $P(S)$ can be estimated by using the relation,
\begin{equation}
P(S)=Z(T)e^{S/T}P_T(S). \label{eq_c4_b6}
\end{equation}
The factor $Z(T)$ is estimated by comparing sampled $P_T(S)$ to 
estimated $P(S)$. {Naturally, this will work only if there is an
overlap in the actually sampled ranges of the support for $P(S)$ from
{unbiased} sampling and $P_T(S)$. 
				This allows already to extend the range
of the support where $P(S)$ is known. Therefore, $P_T(S)$
can be obtained for additional temperatures by starting at a
sufficiently high temperature, allowing to extend
the range of the support even more, again and again, 
where each time $Z(T)$ is calculated
with respect to the so-far known $P(S)$. 
}

Practically, the biased samplings are executed by the 
Markov Chain Monte-Carlo (MC) method.
A graph $G$ is represented by a sequence $\bm{\sigma}$ with 
length $M$ which contains random numbers in $[0,1]$.
{For each pair of nodes, iff the random number is below $p=c/(N-1)$,
an edge connects the two nodes. $S(G)$ {is obtained as} the number of iterations of a simplex
solver which solves Eq.~(\ref{eq_v2}) for $G$. To construct the Markov chain
of the sequences, we used the Metropolis-Hastings
algorithm, which is based on generating trial sequences and accepting
them with a certain probability. To construct a trial
sequence $\bm{\sigma}'$ is copied {from} the current $\bm{\sigma}$ and then 
0.5\% of randomly chosen elements in $\bm{\sigma}'$ are changed.
The corresponding 
 graph $G'$ is generated as explained above. Next,
$S(G')$ is  obtained.}
Finally, $\bm{\sigma}'$ is accepted, {i.e., $\bm{\sigma}'$
becomes the new current sequence in the Markov chain,}
 according to the Metropolis-Hastings acceptance probability which reads
\begin{equation}
P_{\mathrm{acc}}=\max\left\{1,e^{-({S(G')-S(G)})/{T}}\right\}. \label{eq_c4_met}
\end{equation}
{Otherwise $\bm{\sigma}'$ is rejected, i.e., $\bm{\sigma}$ is kept.}

\begin{figure}
\begin{center}
\includegraphics[width=0.97\linewidth]{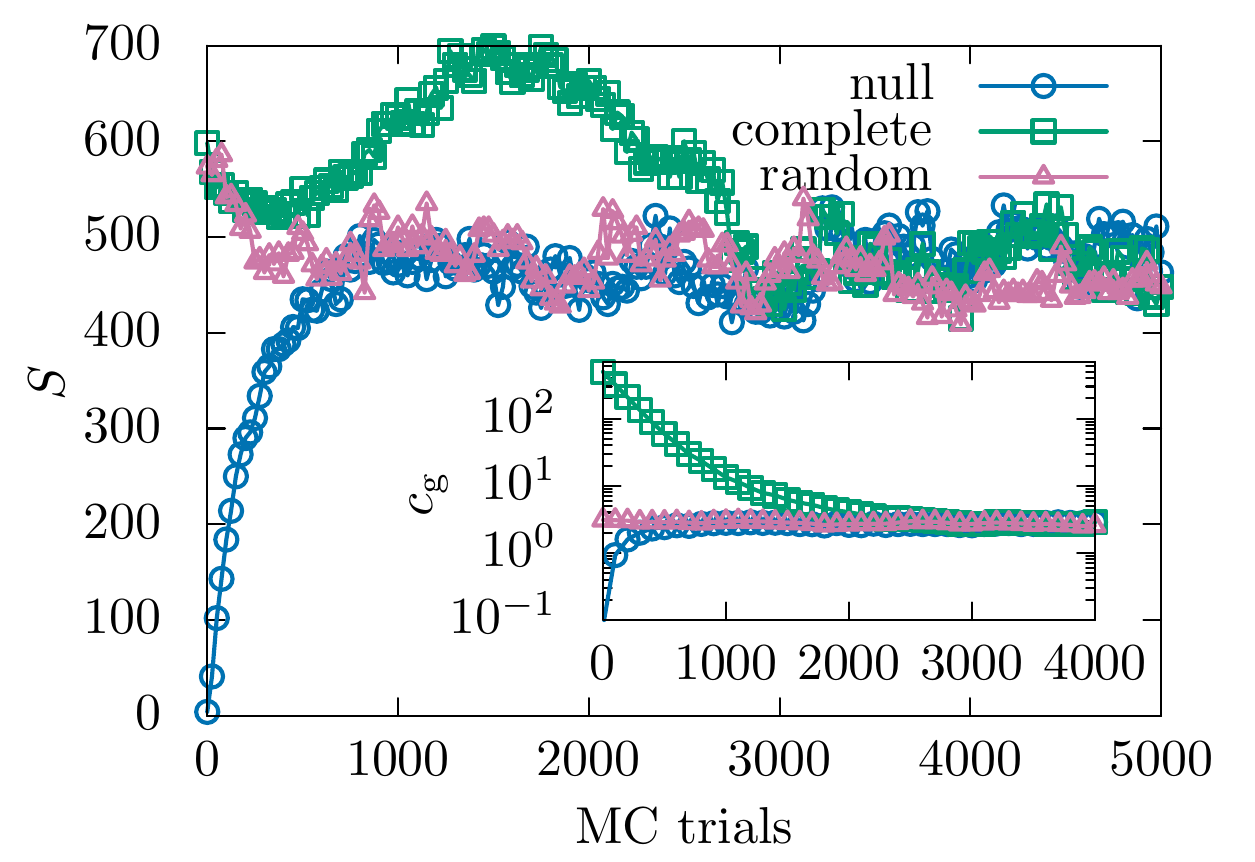}
\caption{(Color Online) An example of the number of LP iterations $S$ 
as a function of the number of MC trials with temperature $T=10$ and 
average degree $c=3$.
Symbols represent numerical results with different initial graphs with $N=500$ vertices; a null graph (circles), a complete graph (squares), and an 
ER random graph {generated from simple sampling} (triangles).
(Inset) Average degree $c_\mathrm{g}$ of a sampled graph as a function of MC trials during in the same simulation.}\label{zu_s1}
\end{center}
\end{figure}

For a biased sampling, one should validate whether the system is equilibrated.
It is a simple way to check sufficient relaxation of observables from different initial conditions.
Fig.~\ref{zu_s1} shows an example of relaxation of two observables, the number of LP iterations $S$ and average degree $c_{\mathrm{g}}$ of each sampled graphs.
In the simulations, the number of vertices $N=500$ and the average degree of ER random graphs $c=3.0$ {is} fixed.
Graphs are sampled according to Eq.~(\ref{eq_c4_met}) with $T=10$ from three different initial graphs, the null graph, the complete graph, and 
a graph simply sampled from ER random graphs.
We find that they 
merge up to 3500 MC trials and then fluctuate similarly around an
equilibrim value 
 indicating that relaxation is promptly realized~\footnote{Usually, the number of updates in the MC sampling is counted by MC steps,
 the number of trials divided by a system size. In this paper, we instead 
use the number of updates itself because of the running time per update and
 the relatively short relaxation time.}.

\begin{figure}
\includegraphics[width=0.97\linewidth]{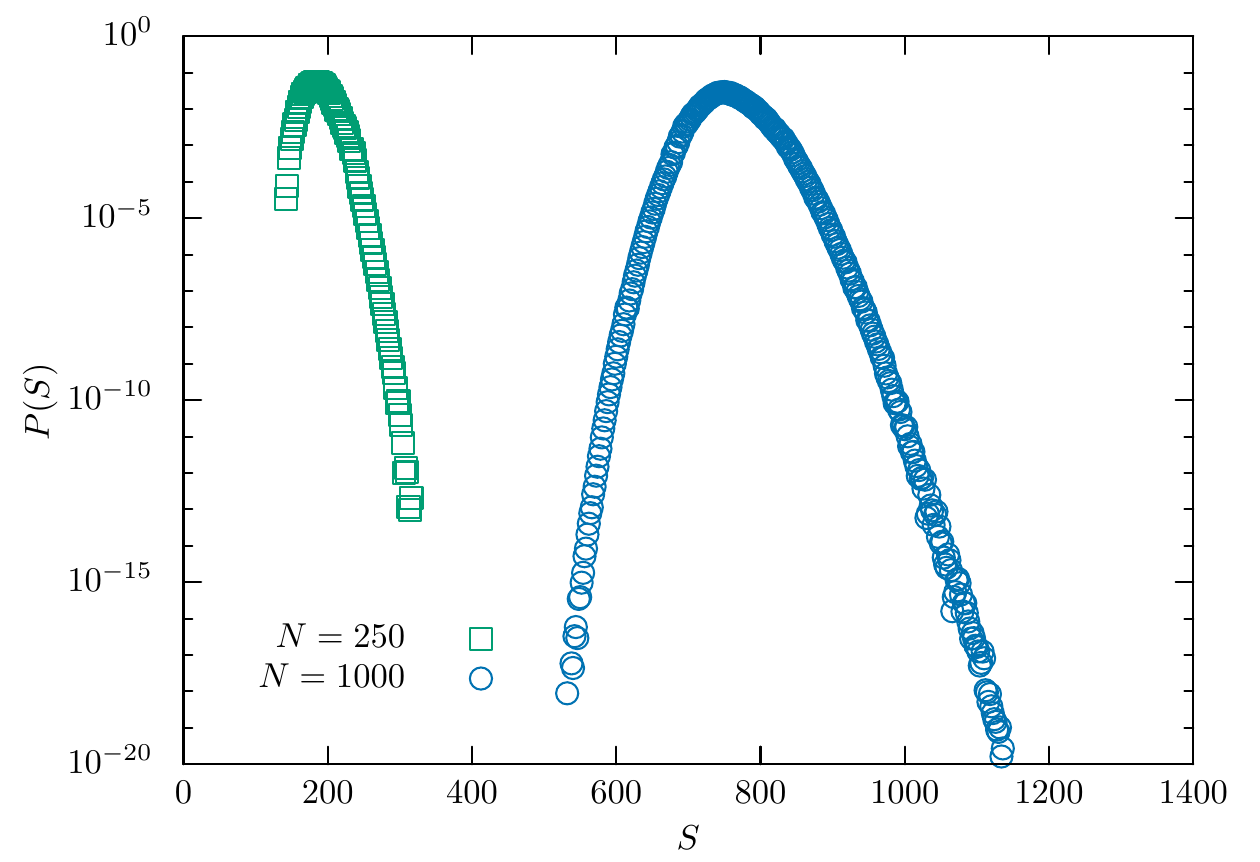}
\caption{(Color Online) Distribution $P(S)$ {of the 
LP iterations $S$ used by lp\_solve for $c=2.0$ and two different graph sizes $N=250$ and $N=1000$.}\label{zu_2}}
\end{figure}

Fig.~\ref{zu_2} shows the estimated  distribution of LP iterations $S$
for ER random graphs with $c=2.0$ for $N=250$ and $1000$.
As a simplex LP solver, lp\_solve is used with default settings.
{Note that we here, for the moment, do  not distinguish between
instances which are {optimally} solved, i.e., all variables are zero or one,
and those instances which are not {optimized}, i.e., where some variables
are half-integer valued.} 
The distribution is generated from an unbiased sampling and biased 
sampling with different temperatures $T=-10,-5,5$, and $10$.
Those samplings are executed for $10^5$ MC trials and the first $2\times 10^4$ observations are omitted to ignore nonequibilium states
for {biased} ones.
The figure shows that the tails of distributions are evaluated up to $10^{-25}$ in spite of executing $10^6$ trials per a sampling.

\section{Results}
In this section, we describe the main results.
First, to evaluate the asymptotic behavior of the  distributions of the number of LP iterations,
an empirical rate function is introduced.
Next, to reveal a relation to the RS-RSB transition, we define hardness of approximation for an instance and
evaluate the distance between conditional  distributions with respect to the hardness.
Its asymptotic behavior suggests a quantitative 
difference of the distance between the RS and RSB phase.
Lastly, {since VC on bipartite graphs can be solved in polynomial 
time, we investigate a quantity}
 defined as the similarity to bipartite graphs {and its relation
to the hardness.}

\subsection{Empirical rate functions}
As indicated in Fig.~\ref{zu_2}, the estimated  distributions themselves are not suitable for comparison.
To visualize their finite-size effects, both axes {should be} rescaled.
For the number of LP iterations $S$, its average $\overline{S}(N)$ over random graphs with $N$ vertices
is used for a linear regression given by
\begin{align}
\overline{S}(N)=S_0+CN^{\alpha}. \label{eq_c4_b7a}
\end{align}
Then, the number $S$ of LP iterations is rescaled by $s=({S}-S_0)N^{-\alpha}$.
In our simulations, the order $\alpha$ is close to one while it has been theoretically unrevealed.

For the distribution $P(S)$, we introduce an empirical rate function which reads
\begin{align}
\Phi(s)=-\frac{1}{N}\ln P(s), \label{eq_c4_b7b}
\end{align}
where $s$ represents the rescaled number of LP iterations.
In the large-deviation theory~\cite{denHollander2009,Tou}, 
{one says that the \emph{large-deviation principle} holds if,
loosely speaking, the empirical rate function
converges for $N\to\infty$ to a limiting rate function
function $\phi(s)$. Thus, in this case it} 
represents the asymptotic behavior of tails of a  distribution as follows:
\begin{align}
P(s)=e^{-N\phi(s)+o(N)}. \label{eq_c4_b7c}
\end{align}
The empirical rate function thus indicates finite-size effects including 
$o(N)$ terms. {Due to the logarithm
and taking $1/N$ to obtain $\Phi(s)$,
the normalization and the subleading term of $P(s)$ become
for finite values of $N$ an additive contribution,
which converges to zero for $N\to \infty$ if the large-deviation principle holds.}

\begin{figure}
\includegraphics[width=0.97\linewidth]{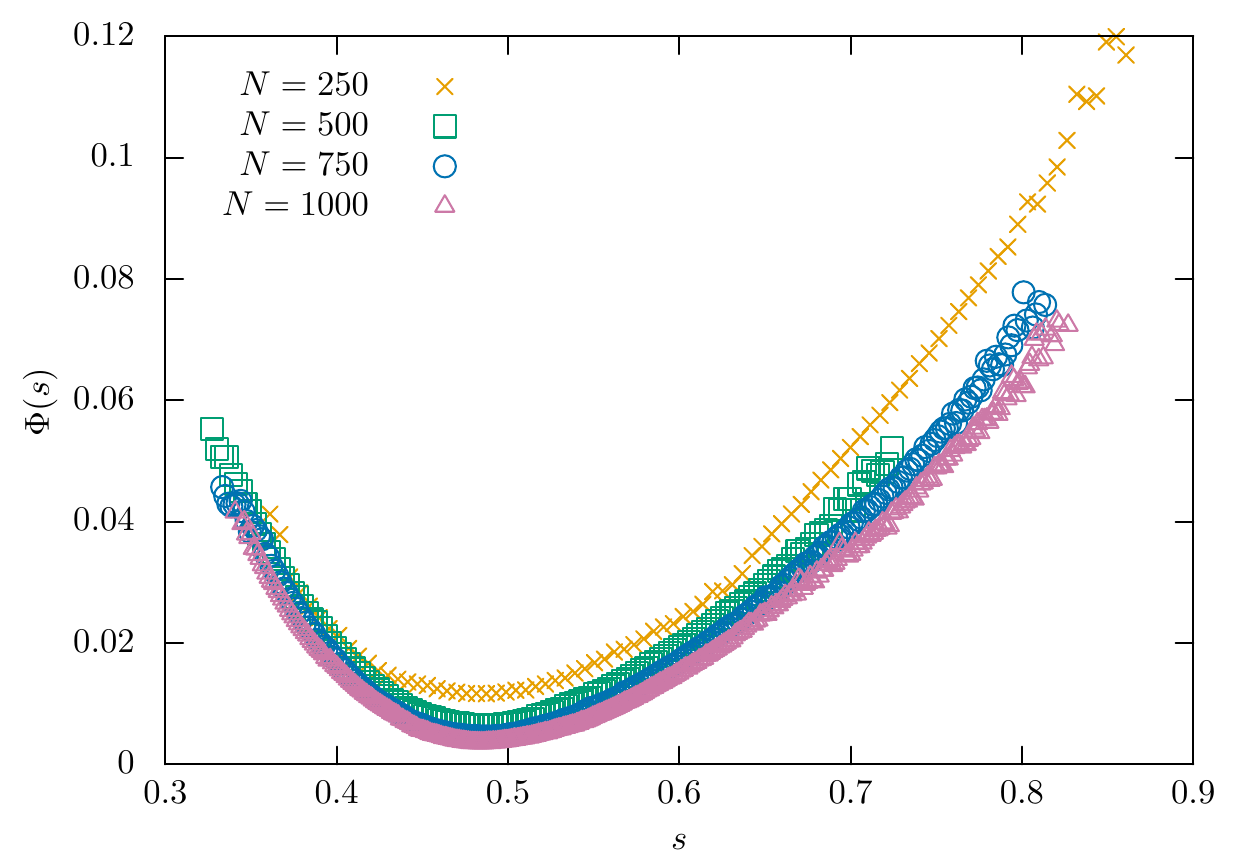}
\caption{(Color Online) Empirical rate functions $\Phi(s)$ for rescaled LP iterations $s$ for LP-relaxed min-VCs on Erd\H{o}s-R\'enyi random graphs with $c=2$
and different cardinalities $N=250$ (cross marks), $500$ (squares), $750$ (circles) and $1000$ (triangles).}\label{zu_s3a}
\end{figure}

\begin{figure}
\includegraphics[width=0.97\linewidth]{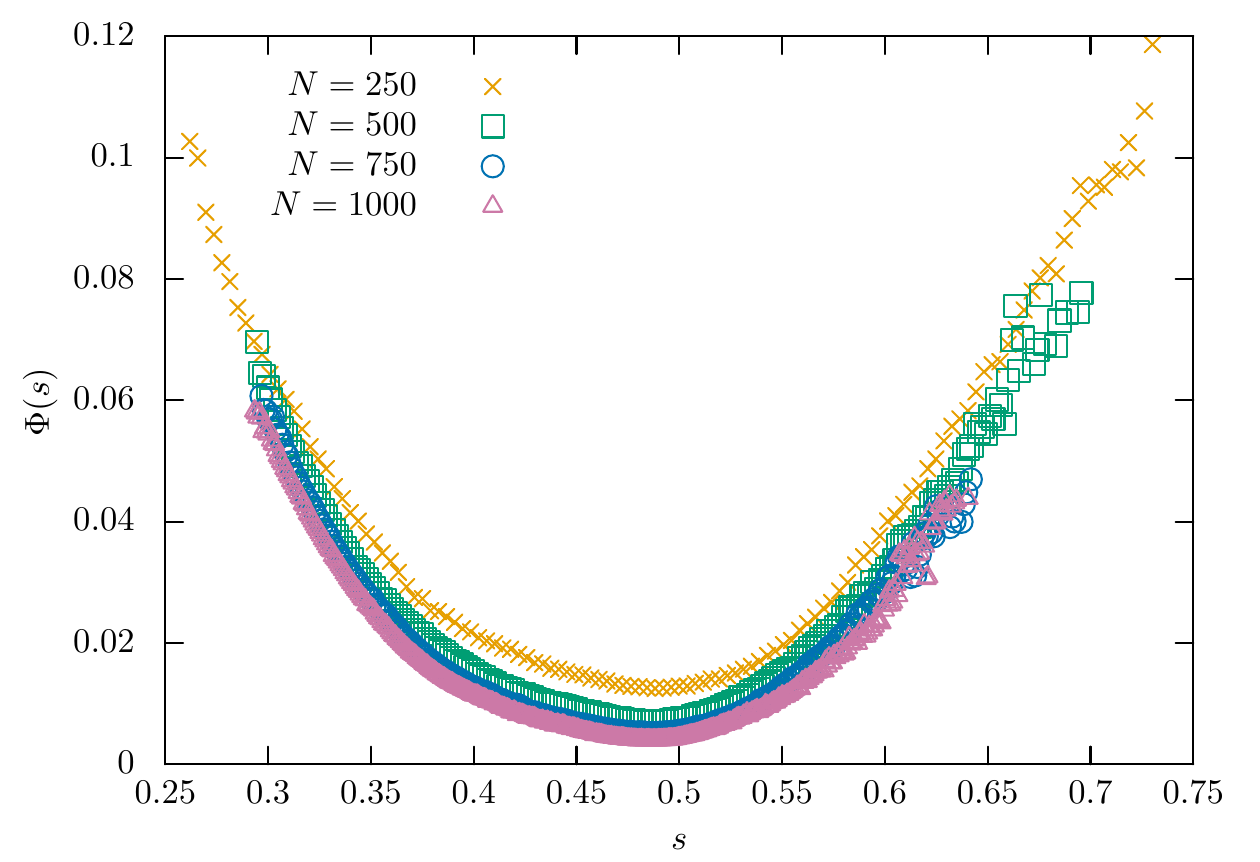}
\caption{(Color Online) Empirical rate functions $\Phi(s)$ for rescaled LP iterations $s$ for LP-relaxed min-VCs on Erd\H{o}s-R\'enyi random graphs with $c=3$
and different cardinalities $N=250$ (cross marks), $500$ (squares), $750$ (circles) and $1000$ (triangles).}\label{zu_s3b}
\end{figure}

Fig.~\ref{zu_s3a} shows the empirical rate functions for ER random graphs with average degree $c=2$, {i.e., in the RS phase,}
 whose  distributions are displayed in Fig.~\ref{zu_2}.
As $N$ grows, the {position on the $s$-axis of the minimum and also the
full shape   of the function  seem to approach to a common position
and a common shape, respectively. This convergence indicates that
the large deviation principle may hold for the rate function of the
number $S$ of LP iterations.}
Since these empirical rate functions are not symmetric with respect to 
their peaks,
 the  distribution of LP iterations differs from a simple Gaussian distribution.
In Fig.~\ref{zu_s3b}, we show empirical rate functions in the case where $c=3$, 
i.e., in the RSB phase.
Compared to Fig.~\ref{zu_s3a}, {a change on} the left-hand side of the 
functions is clearly observed.
This comparison {indicates that the distributions of LP iterations
might exhibit} a qualitative difference related to the RS-RSB phase transition.

\subsection{Distance of  distributions conditioned by hardness of problems}
As described in the last section, a naive expectation is that the computational cost of a simplex LP solver is 
correlated to the hardness of LP relaxation for a problem.
We introduce conditional  distributions of LP iterations with respect to the hardness of problems
to validate the expectation.

For numerical evaluation in this subsection, an alternative LP solver
called CPLEX is executed while the results are qualitatively same as the case of lp\_solve.
The reason is the scalability of CPLEX; it is more than a hundred times as fast as lp\_solve, 
which enables us to set the large number of MC trials  up to $5\times
10^6$ and the large size of graphs up to $3000$.
{We use the rare-event sampling method in the last section with biased
samplings at temperatures $T=-20,-10,5,10$, and $20$ in addition to unbiased samplings.}

\begin{figure}
\includegraphics[width=0.97\linewidth]{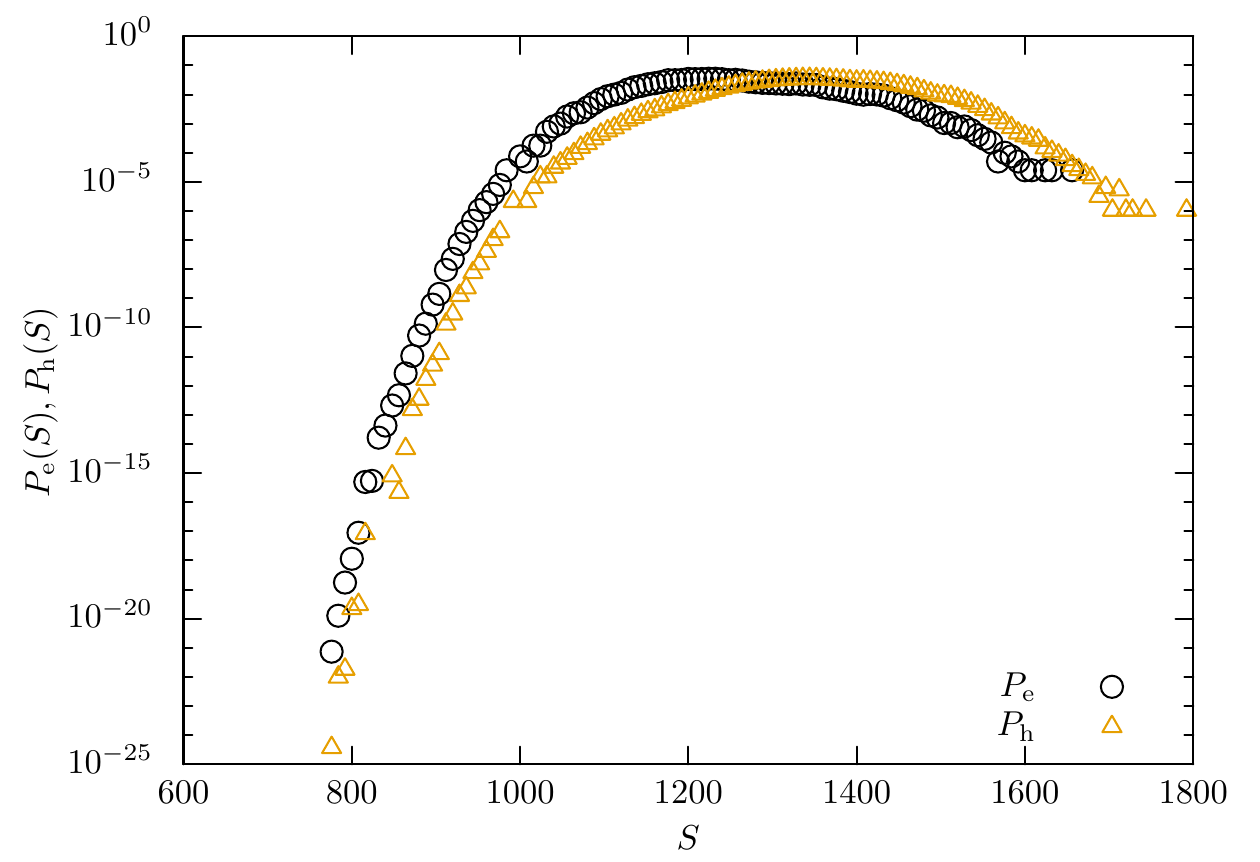}
\caption{(Color Online) Conditional distribution $P_{\mathrm{e}}(S)$ over easy problems (circles) and $P_{\mathrm{h}}(S)$ over hard problems (triangles) 
of LP iteration $S$ by CPLEX for $c=3.0$ and $N=1000$.}\label{zu_c4_6}
\end{figure}

As described in the last section, {an instance} is {considered
 hard to obtain  the optimal solution}  
if the LP-relaxed solution contains at least a  half integer.
{Correspondingly,} conditional  distributions $P_{\mathrm{e}}(S)$ 
and $P_{\mathrm{h}}(S)$ are respectively defined by
 distributions of LP iterations over easy and hard instances, respectively.
Fig.~\ref{zu_c4_6} shows an example of these distributions for ER random graphs with $c=3$ and $N=1000$.
The figure suggests that the expectation about correlation between the number of LP iterations and the hardness of problems 
is correct because two distributions are separated.
The separation is also observed in the case of lp\_solve resulting in the 
{difference} of distributions in Fig.~\ref{zu_s3a} and~\ref{zu_s3b}.

Using conditional distributions, we specifically examine the ``distance'' of those distributions.
Since two sampled distributions generally have different supports, 
the {often used} \emph{Kullback--Leibler distance} 
diverges, for example.
A \emph{total variation distance} is an alternative distance which stays finite 
even in the case of two distributions with completely different supports.
The total variation distance between distributions $P$ and $Q$ is defined by
\begin{equation}
\|P-Q\|_{\mathrm{TV}}=\frac{1}{2}\sum_{x\in I}|P(x)-Q(x)|, \label{eq_c4_c1}
\end{equation}
where $I$ is the union of supports of two distributions.
Unlike the Kullback--Leibler divergence, the total variation distance is a 
kind of metric.
{Since $\sum_{x\in I}P(x)=\sum_{x\in I}Q(x)=1$,} it always 
lies in the interval $[0,1]$ and, especially, takes zero iff $P$ equals to $Q$ 
as a distribution. 

\begin{figure}
\includegraphics[width=0.97\linewidth]{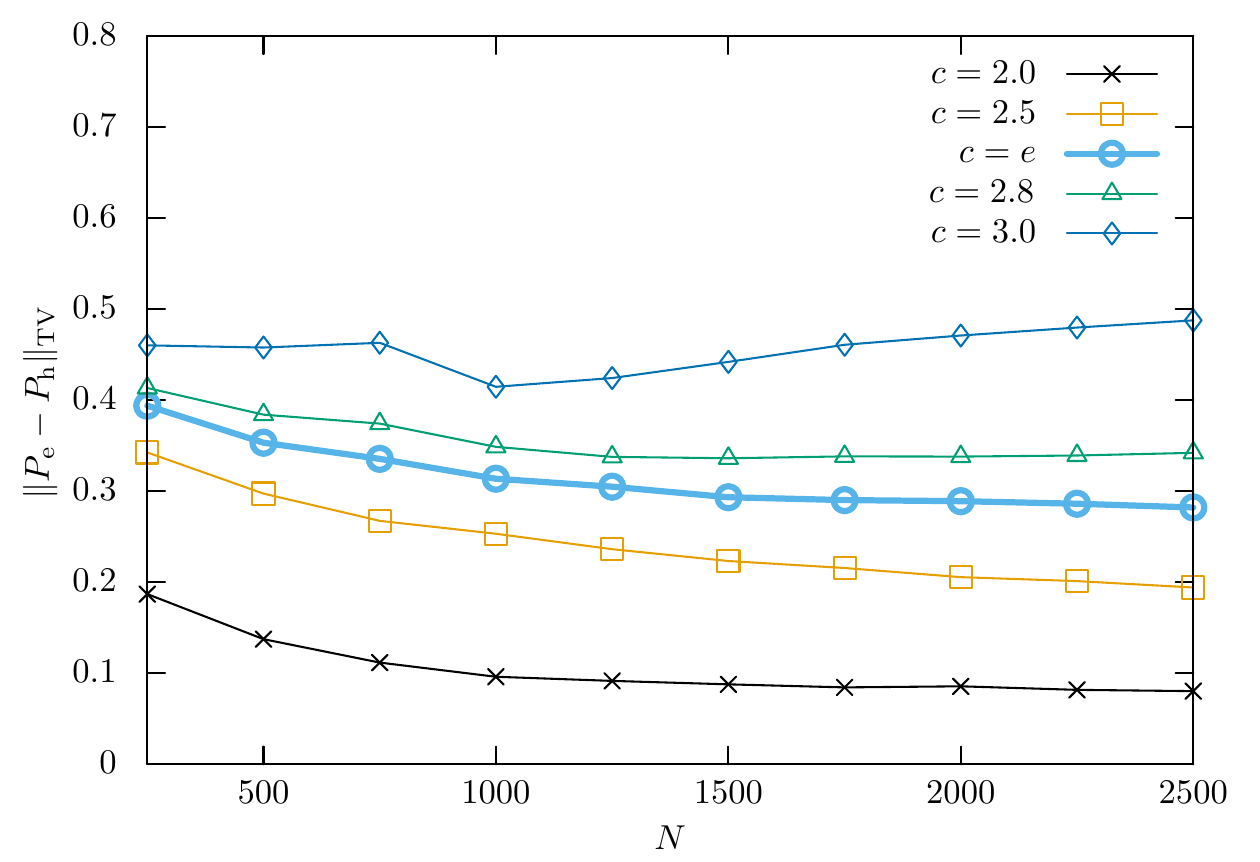}
\caption{(Color Online) Graph size $N$ dependence of total variation
 distance between $P_{\mathrm{e}}(S)$ and $P_{\mathrm{h}}(S)$ with
 different average connectivity $c$.}\label{zu_c4_6a}
\end{figure}

Fig.~\ref{zu_c4_6a} shows the finite-size dependence of the total variation distance between 
conditional distributions $P_{\mathrm{e}}(S)$ and $P_{\mathrm{h}}(S)$ for ER random graphs with various average degree $c\in[2,3]$ {around 
the transition value $c^\ast=e$. Note that for values outside this interval,
it turned out to be hard to obtain sufficient statistics for both
easy and hard instances at the same time, even with the large-deviation
approach.}
In the RS phase where $c<e$, it decreases monotonously and converges nearly to zero.
It implies that {here the} LP computational cost varies {almost} 
independently to the hardness of  LP-relaxed min-VCs in the large-$N$ limit.
On the other hand, for $c>e$, the distance of two distributions 
remains positive even for large $N$ {and seems even to grow, when
looking at the $c=3$ data.}
The fact suggests that the computational cost of the simplex method 
strongly depends on the hardness of instances in the RSB phase.
Asymptotic behaviors of conditional distributions result in the difference of 
 the empirical rate functions in Fig.~\ref{zu_s3a} and Fig.~\ref{zu_s3b}.

\subsection{Relation to graph structure: bipartiteness}
In this subsection, we study a graph invariant {and investigate
{a} possible relation}
to the computational cost of a simplex solver.
{In particular, we introduce ``bipartiteness'' as to which extend 
a given graph
resembles a bipartite graph. A bipartite graph is a graph for which the
set of nodes can be partitioned into two subsets such that there are only
edges which connect one node from one subset to a node from the other
subset. Therefore, there are no edges which connect nodes within one subset
alone.}

In mathematical optimization, K\H{o}nig's theorem is a fundamental result that min-VC problems on bipartite graphs can be solved exactly
in polynomial time~\cite{Kon}.
In ER random graphs, the fact suggests that almost every instances are 
easy to solve with high accuracy
under the percolation threshold, i.e., $c<1$, where a graph consists in trees 
which {are always} bipartite and contain only 
short cycles with length $O(\log N)$. Note that K\H{o}nig's theorem
relates the min-VC to the maximum perfect matching, which can be
found in polynomial time, but we deal with LP here. Nevertheless, it
therefore does not come as a full surprise that LP for the relaxed problem
finds optimal solutions for original min-VC for $c\le 1$.

In contrast, the random graphs {are typically not bipartite}
 above the {percolation} threshold because of the emergence of a loopy giant component.
However, {as we will see below, the still existing 
closeness to bipartite graphs 
allows even for $1<c<e$ to see some correlation between the bipartiteness
and} the number $S$ of LP iterations.

Before describing the numerical results, we define bipartiteness as {the 
fraction of edges in} a maximum cut (max-cut).
A cut of graph $G=(V,E)$ is a partition of vertex set into $T$ and $\overline{T}=V\backslash T$.
Then, a cut set $C(T)$ is defined as a set of edges connecting to each vertex subset $T$ and $\overline{T}$, i.e.,
$C(T)=\{(u,v)\in E;u\in T, v\in\overline{T}\}$.
The maximum cut of $G$ is a cut 
with the largest number of edges among possible cuts.
{The fraction $B$ of edges of a max-cut}
 is then defined as 
\begin{equation}
B=\max_{T\subset V}\frac{|C(T)|}{|E|}\in[0,1]\;.
\end{equation}
It is an indicator of bipartiteness because it equals one iff a graph is 
bipartite.
Since solving a max-cut problem belongs to the class of NP-hard 
{problems, we restricted ourself to calculate
$B$} approximately.
In our numerical simulations, it is approximated by the following greedy 
procedure:
\begin{itemize}
\item[(i)] add each vertex to either of two sets with probability $1/2$, 
\item[(ii)] choose an augmenting vertex named $i$ such that the number of 
neighboring vertices in the same subset is less than
that of neighboring vertices in the other subset, 
\item[(iii)] move $i$ from the current subset to the other subset,
\item[(iv)] if there exists an augmenting vertex, 
return to (iii) or stop otherwise.
\end{itemize}
The algorithm finally finds a locally optimal solution in that it is the best solution
among all cut sets by adding or by deleting a vertex.
{We found that in practice, this algorithm is executed in typically
$10|E|$ iterations
 for} a sampled graph.


\begin{figure}
\includegraphics[width=0.97\linewidth]{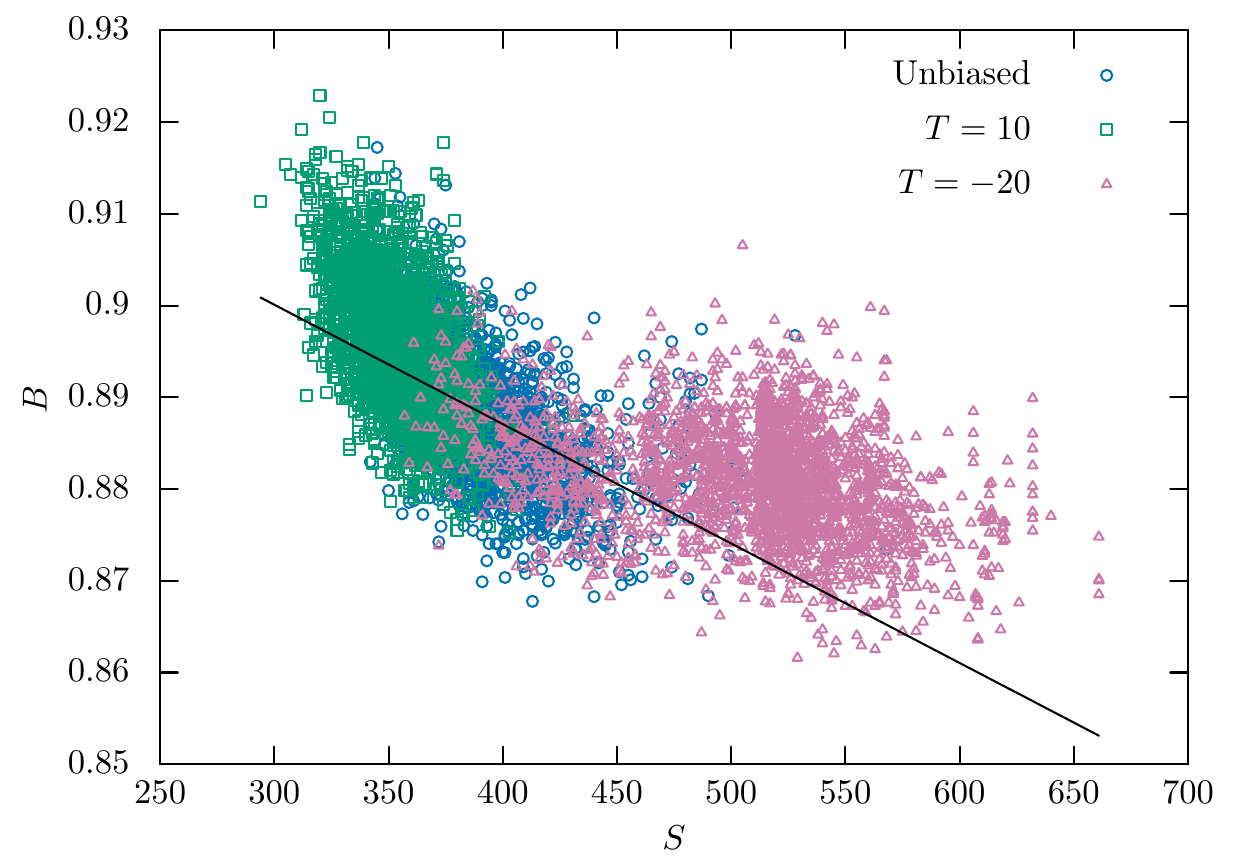}
\caption{(Color Online) Scatter plot of the number of LP iterations $S$ and bipartiteness $B$ by unbiased and biased samplings with different $T$ for $c=2$.
{A solid line represents the result of a linear regression of unbiased samples.}}\label{zu_c4_7a}
\end{figure}

\begin{figure}
\includegraphics[width=0.97\linewidth]{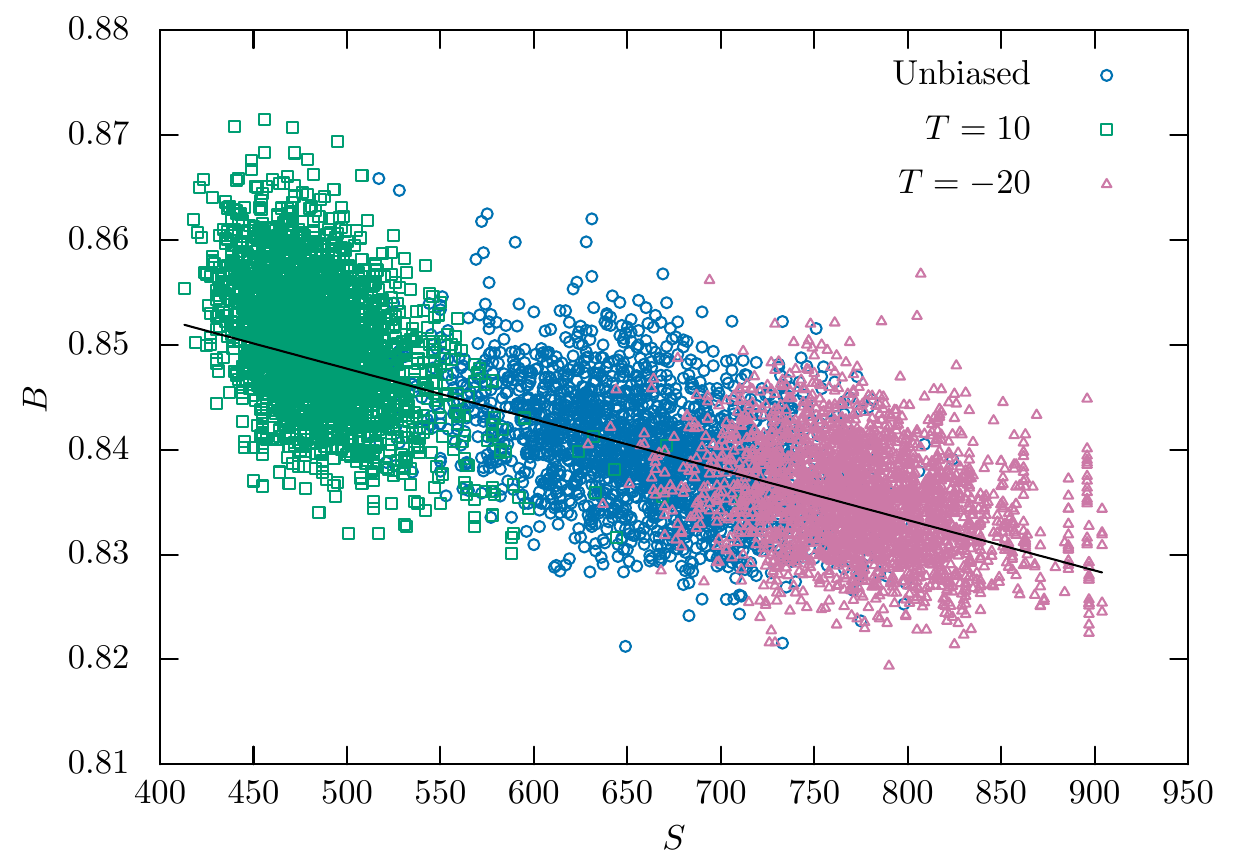}
\caption{(Color Online) Scatter plot of the number of LP iterations $S$ and bipartiteness $B$ by unbiased and biased samplings with different $T$ for $c=3$.
{A solid line represents the result of a linear regression of unbiased samples.}}\label{zu_c4_7b}
\end{figure}

We are interested in the question whether there exists a correlation
between the bipartiteness $B$ and the {number of LP iterations} $S$ to find
a solution.
{
We first show scatter plots of $S$ and $B$ sampled from ER random graphs with $N=500$ and with $c=2$ (Fig.~\ref{zu_c4_7a}) and $c=3$ (Fig.~\ref{zu_c4_7b}).
In those plots, biased samplings with different temperatures $T=-20$ and $10$ are executed in addition to unbiased samplings.
We find that, regardless of RS and RSB phases, $S$ and $B$ have a 
strong {negative} correlation.
However, such a correlation {may result almost trivially 
from a fluctuation of 
the number of edges in ER random graphs: as the number of edges increases,
$S$ also increases because in general graphs with a higher value
of $c$ are harder to solve. On the other hand $B$ should decrease
for a higher number of edges, because
for small values of $c$ almost all graphs are collections of trees, which
are completely bipartite, while densely connected graphs are not bipartite.}
To omit such a spurious correlation, it is necessary {to consider an
ensemble of random graphs, where} the number of edges is fixed to $M=cN/2$.
In the thermodynamic limit $N\to\infty$, both ensembles agree 
\cite{erdoes1960}}.

\begin{figure}
\includegraphics[width=0.97\linewidth]{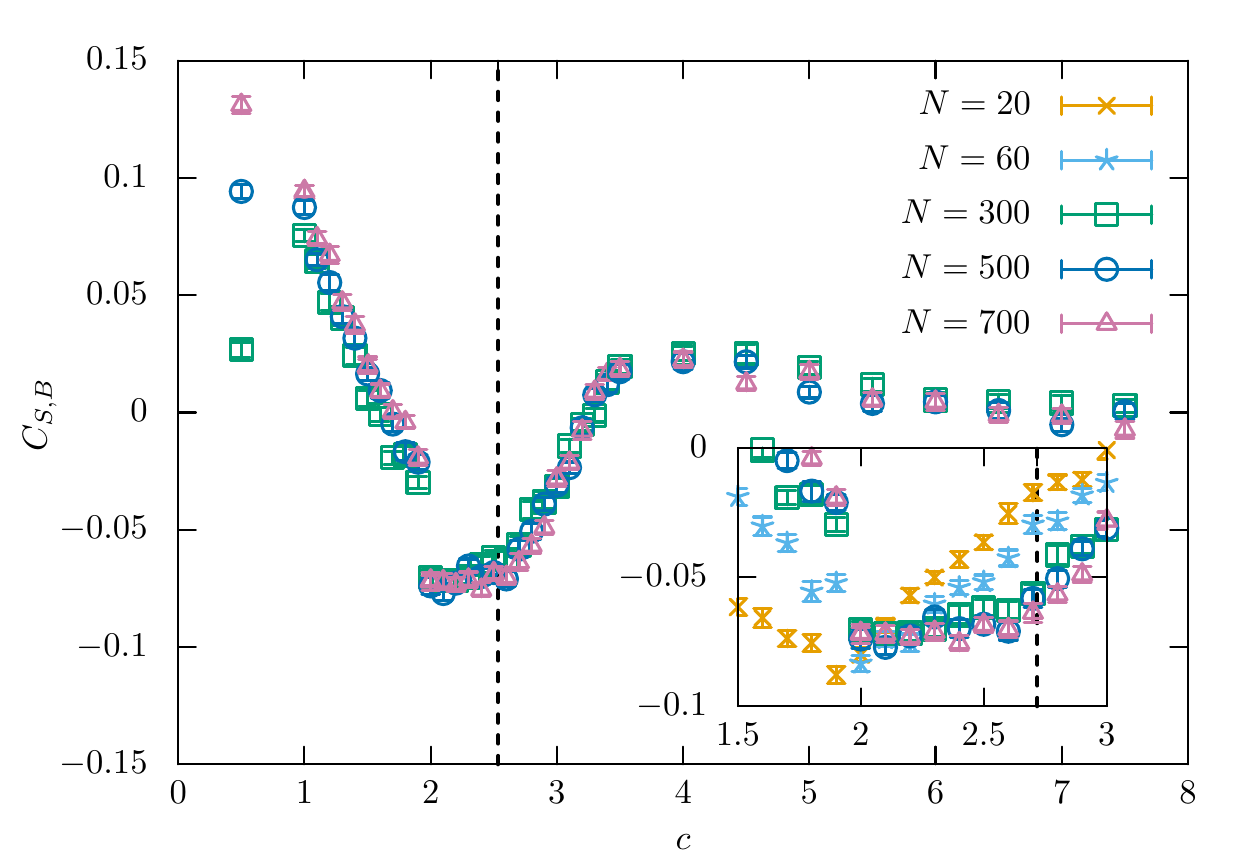}
\caption{(Color Online) Correlation coefficient $C_{S,B}$ between $S$ and $B$ as a function of average degree $c$.
Each symbol represents a numerical estimation of $C_{S,B}$ obtained by unbiased samplings with different numbers of vertices;
 $N=20$, $60$, $300$, $500$, and $700$. 
The vertical line is the RS--RSB threshold given by $c^\ast=e$.
{(Inset) An enlarged plot around the RS--RSB threshold.}}\label{zu_c4_8}
\end{figure}

{
Now we concentrate on the estimation of correlation between $S$ and $B$ for ER random graphs with the number of edges fixed.
The correlation is calculated as the Pearson correlation coefficient defined by
\begin{equation}
C_{S,B}=\frac{\overline{(S-\overline{S})(B-\overline{B})}}{\sqrt{\overline{(S-\overline{S})^2}}\sqrt{\overline{(B-\overline{B})^2}}}, \label{eq_d1}
\end{equation}
where $\overline{\cdots}$ represents an average over sampled graphs.
Fig.~\ref{zu_c4_8} shows the correlation coefficient $C_{S,B}$}
estimated by unbiased samplings with $10^5$ graphs whose number of edges is fixed to $cN/2$. 
CPLEX is used as a {simplex} solver again.
{The standard deviation of $C_{S,B}$ is estimated by the bootstrap method.
Namely, the data are uniformly resampled and a correspondent $C_{S,B}$ is calculated for 50 times to estimate its standard deviation.
As shown in Fig.~\ref{zu_c4_8}, it is found that the 
finite size effect looks small {for $N\ge 300$
 above the percolation threshold $c=1$, since the data points fall 
on top of each other for different values of $N$.}
In the case of $c=0.5$, the sampled data shapes just a point in the scatter plot because all sampled graphs are nearly bipartite.
It makes an accurate estimation of $C_{S,B}$ difficult.
Above the percolation threshold, $C_{S,B}$ gradually decreases and becomes negative above $c\sim 1.7$.
The gap of $C_{S,B}$ between $c=1.9$ and $2$ emerges because of an intrinsic behavior of the solver.
It must be emphasized, however, that relatively strong negative correlations are observed below the RS--RSB threshold and 
the correlation moves nearly to zero above the threshold regardless of a simplex solver.
}

{In summary, even though a trivial correlation related to the number of edges is omitted, 
we find a relatively weak but characteristic behavior of the correlation coefficient between the number of LP iteration $S$ and bipartiteness $B$.}
Consequently, the intuition that similarity to bipartite graphs decreases 
the computational cost of a simplex LP solver
is {(slightly)} correct only in the RS phase where the majority of 
graphs in the ensemble is easy to obtain optimal solutions.
{Note that we have studied also other graph invariants such as a 
Becchi number and {degree correlations} to seek for a correlation to the number of
LP iterations. So far, we did not find such a correlation for any value of $c$.}

\section{Summary and Discussions}
In this paper, we numerically investigate the {large-deviation properties} of computational cost for LP relaxation
{and relations} to a phase transition in the spin-glass theory and to random graph structure.
The rare-event sampling is executed to {efficiently examine} the large-deviation properties.
It enables us to estimate the tails of the distributions efficiently illustrating its asymmetric property and finite-size effects.
Furthermore, we naively introduce the hardness of LP relaxation and evaluate the total variation distance between conditional distributions with 
respect to the hardness.
Numerical results show that the distance asymptotically {becomes
very small} for $c<e$ while it remains {larger and seems even to grow
with graph size} otherwise.
It is indicated that conditional distributions are asymptotically distinguishable in the RSB phase, which reflects the hardness of {instances for} LP relaxation.
The results are compatible to the {differences observed} in the empirical rate functions.
Finally, the relation of computational cost of LP relaxation to graph structure is studied.
We specifically examine bipartiteness of graphs defined as the 
fraction {of edges in the} maximum cut.
It is suggested that it is weakly but negatively correlated to the number of LP iterations in the RS phase
though they are nearly uncorrelated otherwise.

The rare-event sampling based on the reweighting method helps us to find a 
valid {signature} in the distribution $P(S)$ in this paper.
To access the further details of the tail of the distribution, a 
biased sampling with lower temperature {can be performed
in principle.}
It is observed, however, that the system with negatively low temperature, 
e.g., $T=-1$, is hard to equilibrate using
 the Metropolis-Hastings algorithm.
It {could be beneficial} to use another rare-event sampling 
scheme such as the Wang-Landau method or an advanced MC sampling method like
the replica exchange MC method~\cite{91}.

As for the relation to graph structure, we find a valid negative correlation to the computational cost of LP relaxation in the RS phase
though it vanishes in the RSB phase.
A possible reason of the uncorrelated result for $c>e$ is the hardness of estimating the bipartiteness in the region.
In fact, the max-cut problem on the ER random graphs corresponding to finding the ground energy of the antiferromagnetic spin-glass model~\cite{ZB}
also exhibits the RS-RSB transition at average degree $c=e$.
It suggests that the greedy procedure in the last subsection fails to find the optimal fraction of max-cut even though it is executed repeatedly.
While the large part of graph invariants is hard to compute exactly, 
{we have to leave it to} a future work to find {even better} 
one to characterize the complexity of graph structure even in the RSB phase.

In the view of the spin-glass theory, it is {considered} that local structures called frustrations evoke the RSB phase~\cite{26}.
For Ising spin glasses on a finite dimensional lattice, for example, there is an attempt to detect the transition point using
frustrated plaquettes, i.e., frustrated shortest loops in the lattice~\cite{Miya,Ohz}.
In the case of a spin-glass model on a sparse random graph, however, it is not straightforward to define such a local structure
because graph invariants describe rather global structure.
A long range frustration~\cite{Zhou1} based on the cavity method is a strong candidate though it is difficult to evaluate numerically in practice.
It is thus an important task to find another measure of frustrations in random graphs to predict the hardness of each instance.
The result in this paper suggests that the LP relaxation is possibly useful to investigate the measure numerically by evaluating the correlation
to the computational cost of a solver.

Evaluating the typical computational cost of a LP solver over randomized problems {by analytical approaches} is highly challenging
though the {worst-case} computational cost has been a central 
issue of mathematical optimization.
Our study on the large-deviation properties shows the existence of the common properties irrelevant to the details of
solvers and of nontrivial relations to the phase transition, the spin-glass picture, and random graph structures.
We believe that numerical analyses in this paper stimulate the further progress in understanding typical behavior of LP relaxation
for randomized optimization problems and its vast relationships to other concepts including the spin-glass theory.

\begin{acknowledgments}
The use of IBM ILOG CPLEX has been supported by the IBM Academic Initiative.
This research is partially supported by JSPS KAKENHI Grant Nos.~25120010 (KH) 
and~15J09001 (ST).
\end{acknowledgments}


\begin{thebibliography}{99}
\bibitem{phase-transitions2005} A. K.~Hartmann and M.~Weigt,
  \textit{Phase Transitions in Combinatorial Optimization Problems},
  (Wiley-VCH, Weinheim ,2005).

\bibitem{26} M.~M\'ezard and A.~Montanari, 
\textit{Information, Physics, and Computation}
 (Oxford University Press, Oxford, 2009).

\bibitem{moore2011} C.~Moore and S.~Mertens,
  \textit{The Nature of Computation},
  (Oxford University Press, Oxford, 2011).

\bibitem{papadimitriou1998} C.~Papadimitriou and K.~Steiglitz,
\textit{Combinatorial Optimization -- Algorithms and Complexity},
(Dover Publications Inc., Mineola, NY, 1998).

\bibitem{garey1979} M. R.~Garey and D. S.~Johnson, 
\textit{Computers and Intractability}, (Freeman, San Francisco, 1979).

\bibitem{mitchell1992} D.~Mitchell, B.~Selman, and H.~Levesque,
in Proceedings of the AAAI-92, 459 (1992).

\bibitem{monasson1997} R.~Monasson and R.~Zecchina, 
Phys. Rev. E \textbf{56}, 1357 (1997).

\bibitem{mertens1998} S. Mertens, Phys. Rev. Lett. \textbf{81}, 4281 (1998).

\bibitem{wh1} M.~Weigt~and A.~K.~Hartmann, 
Phys. Rev. Lett. \textbf{84}, 6118 (2000).

\bibitem{horie1996} A.~Horie and O.~Watanabe, 
Lect. Notes Comput. Sci. \textbf{1350}, 22 (1996).

\bibitem{hoos2000} H.~Hoos and T.~St\"utzle, 
J. Autom. Reasoning \textbf{24}, 421 (2000).

\bibitem{liu2015} R.~Liu, W.~Luo, and L.~Yue,
Data Knowl. Engin. \textbf{100}, 1-18 (2015).

\bibitem{jia2007} H.~Jia, C.~Moore and D.~Strain,
J. Artif. Intell. Res. \textbf{28}, 107-118 (2007).

\bibitem{satsat2002} W.~Barthel, A. K.~Hartmann, M.~Leone, F.~Ricci-Tersenghi,
M.~Weigt, and R.~Zecchina, Phys. Rev. Lett. \textbf{88}, 188701 (2002).

\bibitem{zdeborova2006} L. Zdeborova and Marc M\'ezard, 
J. Stat. Mech.\textbf{2006}, P12004 (2006).

\bibitem{marshall2016} J.~Marshall, V.~Martin-Mayor, and I. Hen,
Phys. Rev. A \textbf{94}, 1 (2016).

\bibitem{AKH2002} A. K. Hartmann, Phys. Rev. E, \textbf{65}, 056102 (2002).

\bibitem{AKH2011} A. K. Hartmann, Eur. Phys. J. B, \textbf{84}, 627 (2011).

\bibitem{fBm_MC2013} A. K.~Hartmann, S. N.~Majumdar, and A. Rosso,
  {Phys. Rev. E}, \textbf{88}, 022119 (2013).

\bibitem{work_ising2014} A. K.~Hartmann,
  {Phys. Rev. E} \textbf{89}, 052103 (2014).

\bibitem{erdoes1960} P.~Erd\H{o}s and A.~R\'enyi,
Publ. Math. Inst. Hungar. Acad. Sci. \textbf{5}, 17--61 (1960).

\bibitem{Tak} J. Takahashi, S. Takabe, and K. Hukushima, 
J. Phys. Soc. Jpn. \textbf{86}, 073001 (2017).

\bibitem{dh} T.~Dewenter~and A.~K.~Hartmann, 
Phys. Rev. E \textbf{86}, 041128 (2012).

\bibitem{Kha} L. G. Khachiyan, USSR Comput. Math. Math. Phys. 20, 53 (1980). 
\bibitem{Kar} N. Karmarkar, Proc. of the sixteenth annual ACM symp. on Theo. of comp. 302 (1984).
\bibitem{nt} G.~L.~Nemhauser and L.~E.~Trotter Jr., Math. Program. \textbf{6}, 48 (1974).
\bibitem{th2} S.~Takabe and K.~Hukushima, J. Phys. Soc. Jpn. \textbf{83}, 043801 (2014).
\bibitem{th4} S. Takabe and K. Hukushima, J. Stat. Mech. \textbf{2016}, 113401 (2016).
\bibitem{WL} F. Wang and D. P. Landau, Phys. Rev. Lett. \textbf{86}, 2050 (2001).
\bibitem{berg} B. A. Berg and T. Celik: Phys. Rev. Lett. \textbf{69}, 2292 (1992).
\bibitem{IH} Y. Iba and K. Hukushima, J. Phys. Soc. Jpn. \textbf{77}, 103801 (2008).
\bibitem{AKH2015} T. Dewenter and A. K. Hartmann, 
New J. Phys. \textbf{17}, 015005 (2015).
\bibitem{AKH2017} A. K. Hartmann, Eur. Phys. J. Spec. Top. \textbf{226}, 
567 (2017). 
\bibitem{AKH2017_2} A. K. Hartmann and M. M\'ezard, arXiv:1710.05680 (2017).
\bibitem{practical_guide2015} A. K. Hartmann,
\textit{Big Practical Guide to Computer Simulations},
(World Scientific, Singapore, 2015)

\bibitem{denHollander2009} F. den Hollander, \textit{Large Deviations},
  (American Mathematical Society, Providence, 2000)
\bibitem{Tou}  H. Touchette, Phys. Rep., \textbf{478}, 1 (2009).
\bibitem{Kon} D. K\"onig, Matematikai \'es Fizikai Lapok, \textbf{38}, 116 (1931).
\bibitem{91} K.~Hukushima and K.~Nemoto, J. Phys. Soc. Jpn. \textbf{65}, 
1604 (1996).
\bibitem{ZB} L. Zdeborov\'a and S. Boettcher, J. Stat. Mech. \textbf{2010}, 
02020 (2010).
\bibitem{Miya} R. Miyazaki, J. Phys. Soc. Jpn. \textbf{82}, 094001 (2013).
\bibitem{Ohz} M. Ohzeki, Y. Kudo, and K. Tanaka, J. Phys. Soc. Jpn. \textbf{87}, 015001 (2017).
\bibitem{Zhou1} H. Zhou, Phys. Rev. Lett. \textbf{94}, 217203 (2005).

\end{thebibliography}
\end{document}